# Spacetime G-structures I: Topological Defects

D.H. Delphenich[†]
Physics Department
University of the Ozarks
Clarksville, AR 72840

*Abstract. The notion of G-structure is defined and various geometric and topological aspects of such structures are discussed. A particular chain of subgroups in the affine group for Minkowski space is chosen and the canonical geometrical and topological objects that are associated with each reduction of the bundle of affine frames on a four-dimensional manifold are detailed. Their physical significance is discussed in the language of topological defects in ordered media. Particular attention is paid to how one topologically characterizes the wave phase of the spacetime vacuum manifold.*

**0. Introduction.** In the usual theory of spontaneous symmetry breaking in ordered media [**1-7**], one is generally concerned with pairs of subgroups in some symmetry group G ⊃ H, their homogeneous spaces G/H, and the homotopy groups $\pi_\bullet$(G/H). The group G represents some symmetry in the states of the medium and the subgroup H is referred to as the *hidden, spontaneously broken, or residual symmetry* in the system, which is generally assumed to have a G symmetry. One refers to the homogeneous space G/H as either the *space of order parameters* or the *vacuum manifold* for the ordered medium. Often, one also looks at chains of subgroups G ⊃ H ⊃ K ⊃ …, and in such cases the sequence of homogeneous spaces G/H, H/K, …, represents *phases* of the ordered medium, since the spontaneous breaking of an internal symmetry is generally associated with a physical phase transition such as a transition from a disordered phase to a magnetically polarized, superconducting, or superfluid phase.

The *order parameter* for the ordered medium M is a map from M to G/H. One frequently looks at homotopy classes [M; G/H] of such maps, which represent the stable configurations of the order parameter. In the event that the manifold M is itself a sphere of some dimension – say $k$ – these homotopy classes become elements of $\pi_k$(G/H). When this group is non-vanishing, one says that one has encountered a *topological defect* in dimension $k$. By now, a consistent vocabulary of such items has settled out, and a non-vanishing element of $\pi_0$(G/H) is called a *wall,* such an element of $\pi_1$(G/H) is called a *string* (or *vortex*). In $\pi_2$(G/H), one speaks of *monopoles,* and in $\pi_3$(G/H), one encounters *textures.*

The objective of this study is to pursue this sort of ordered medium analogy in the context of the bundle of linear frames on spacetime, which is considered to be a connected four-dimensional manifold M. (The role of a Lorentz structure will be discussed as a stage in the breaking of the symmetry of the spacetime vacuum manifold, which is not to be confused with the spacetime manifold itself.) The generalization is that instead of looking at homotopy classes of maps [M; G/H] we shall be more concerned with homotopy classes of maps [M; G/H(M)], in which G/H(M) is a homogeneous bundle over M, i.e., a fiber bundle over M whose fibers are diffeomorphic to the homogeneous space G/H. Such bundles will appear in the process of reducing from a G-structure to an H-structure on M, where a G-structure is a reduction of GL(M), the bundle of linear frames on M to a bundle of frames that defines a G-principal bundle on M, and G is a subgroup of GL($n$). The aforementioned simpler case of [M; G/H] that appears in condensed matter is included in this picture under the assumption that G/H(M) is trivial, an assumption that will not be made in the present article. Since a section of the bundle G/H(M) → M is equivalent to a G-equivariant map $\phi$: G(M) → G/H, triviality would imply the existence of a global section of G(M), i.e., a global G-frame field **e**: M → G(M). One could then compose $\phi$ with **e** to define a map $\phi$·**e**: M → G/H.

---

[†] ddelphen@ozarks.edu

However, the parallelizability of manifolds is such a topologically sensitive issue that we shall assume that this situation is only locally possible, and examine the topological obstructions to the existence of global sections of both GL(M) and G/H(M). The obstructions are cocycles in the cohomology of M that take their values in the homotopy groups of G/H. This is also essentially a generalization of the condensed matter scenario since a cohomology class is, after all, a homotopy class of maps, but its domain is only some particular chain in M and not all of M itself. This is also a convenient mechanism for a discussion of localized symmetry-breaking processes, which are, of course, more physically meaningful than the ones in which ferromagnets, liquid crystals, or superfluids are allowed to fill the known universe. This consideration is also unavoidable in the eyes of topology, which is indifferent to the orders of magnitude that distinguish the subatomic scale from the cosmological scale. Furthermore, one is no longer limited to spacetime manifolds that take the form of 4-spheres or the product of a real (time) line with a 3-sphere, since the only spheres in M that contribute to the various groups $\pi_k(G/H_x)$ are the ones that one finds in the cells that one uses to represent M as a CW-complex.

We shall consider a particular sequence of reductions GL(M) ⊃ $G_1$(M) ⊃ $G_2$(M) ⊃ … ⊃ {$e$}(M) that is defined by a chain of subgroups, GL(4) ⊃ $G_1$ ⊃ $G_2$ … ⊃ {$e$}, and look at the topological obstruction to the reduction from each $G_i$(M) to $G_{i+1}$(M), which will be an element of H*(M; $\pi_*(G_i/G_{i+1})$). In some cases, the obstruction cocycle takes the form of commonly known characteristic classes on M, such as the Euler or Stiefel-Whitney classes.

We shall then discuss the relationship between obstructions and the singularity complex of the chain of reductions, and how it generalizes the usual picture of topological defects. We shall investigate the extent to which the relationship of the fundamental tensor field of a reduction (if such a field exists) to the associated topological defect is that of a field to its "source."

We will devote especial attention to the chain of reductions and obstructions that lead to a spacelike SO(2)-structure on spacetime, which was elsewhere **[8, 9]** shown to be the basic structure that one needs in a four-dimensional manifold in order to facilitate wave motion, independently of one's choice of wave equation.

In a subsequent study, we shall also look at the matter of whether a connection defined on a G-structure is reducible to a connection that is defined on an H-structure (H⊂G), and try to give a continuum mechanical interpretation for the part of the connection that does not reduce as a measure of the deformation of the H-structure from its "ground state" in that phase.

**1. G-structures [10-14]**.  Let M be an *n*-dimensional differentiable manifold and let us denote its bundle of linear frames by GL(M); this bundle is a principal GL(*n*)-bundle over M.  If G is a subgroup of GL(*n*) then one calls any reduction of GL(M) to a G-principal bundle, which we generally denote by G(M), a G-*structure* on M.  Here, an H-*reduction* of a G-principal bundle P is a submanifold of P that defines an H-principal bundle over M along with the restriction of the projection of GL(M) on M.

Examples of G-structures in differential geometry are so numerous and important, one must conjecture whether "everything in differential geometry is a G-structure (a.e.)"  For example, an orientation of the manifold M is associated with a reduction of the linear frames to oriented linear frames, hence, a reduction from GL(M) to $GL^+(M)$.  A Riemannian metric on M is equivalent to a reduction of GL(M) to O(M), that is reducing from linear frames to orthonormal frames.  Naturally, one can further reduce to SO(M), i.e., oriented orthonormal frames.  Similarly, a Lorentz structure on M is a reduction from GL(M) to O(1,3)(M), i.e., Lorentz orthonormal frames.  Ultimately, one might reduce to an {*e*}-structure, which is a unique choice of frame at each point of M, i.e., a global frame field on M.  As this example clearly shows, reductions to subgroups are not always possible and depend upon the topology of M; we shall return to this subject shortly.  Some other geometric structures on manifolds that can be represented by G-structures are differential systems, symplectic structures, and almost-complex structures, although we shall not make further use of these facts in the present work.

It can be shown [**15**] that the existence of a reduction of a G-principle bundle P → M to an H-principle bundle is equivalent to the existence of a global section of the *associated homogeneous bundle* that is defined by P and G/H.  This is a fiber bundle over M with fiber G/H and structure group G that one obtains by looking at the orbits of the action of G on P×G/H by way of $g(p, v) = (pg^{-1}, gv)$, and one denotes this bundle by P×$_G$G/H, or, when no confusion will arise, G/H(M)  More directly, one can consider the action of H on GL(M) and think of a point of a fiber of this bundle at any *x*∈M as representing an orbit of the action of H on $GL_x(M)$.

These sections, in turn, are in one-to-one correspondence with the G-equivariant maps from P to G/H.  Since the principal bundles that we are dealing with will be frame bundles over differentiable manifolds and the homogeneous spaces G/H will be equivalence classes of real invertible *n*×*n* matrices, there will be instances when these equivariant maps take the form of tensor fields on M, which we call the *fundamental tensor field* of the G-structure that we have defined.  In all cases that we will consider in the present study, these tensor fields will be smooth functions, vector fields, or second rank tensor fields on M.

For the sake of examining the topological obstructions to the reduction of a G-structure to an H-structure, it is more convenient to use the characterization of a reduction from G(M) to H(M) as a global section of the associated fiber bundle G/H(M) = G(M)$\times_G$G/H. This is then a problem in the standard form for the application of obstruction theory [**16, 17**].

The basic process in determining the obstructions to global sections of a fiber bundle B → M (or just global continuous maps between topological spaces, in general) is to express the base manifold M as a CW-complex ([1]) |M| (up to homotopy equivalence), and then:

*a)*  Start with the 0-skeleton of |M|, define a section $\phi_0$ over it,

*b)*  Look at the $0^{th}$ homotopy group of the fiber F at each point of the 0-skeleton to see if it vanishes, i.e., see if F is path-connected,

*c)*  If $\pi_0(F_x) = 0$ then extend $\phi_0$ to a section $\phi_1$ on the 1-skeleton by a path in F, etc.

Eventually – or not – one reaches a dimension $k-1$ in which $\pi_{k-1}(F) \neq 0$, so one cannot extend $\phi_{k-1}$ to the cells of the $k$-skeleton of |M| by homotopy. The *primary obstruction* to the existence of such a global section then takes the form of a cohomology class:

$$\mathfrak{o}_k(B) \in H^k(M, \pi_{k-1}(F)),$$

which expresses the obstruction to extending a section that is defined over the $k-1$-skeleton of |M| to a section that is defined over the $k$-skeleton. Note well that the vanishing of such an obstruction is a necessary and sufficient condition for the existence of the extension *to the k-skeleton*, but it is *not sufficient* to insure the existence of a *global* section of the bundle in question. Note also that the homotopy group at some stage may be non-trivial, but the cocycle can still vanish for reasons related to the topology of M itself.

The coefficient sheaf consists of the sheaf of all $(k-1)^{th}$ homotopy groups of F, when F is represented as the fibers of B. This means, in particular, that although the groups at any two points on the same path component of M will be isomorphic by conjugation, nevertheless, since the isomorphism is not canonical the sheaf is not necessarily constant; that is, there might be a relative "twisting" of distinct fibers.

Once one has established the first non-trivial homotopy group of the fibers, $\pi_{k-1}(F_x)$, one can also get information about how the sections of B → M are partitioned into homotopy equivalence classes. Go back to the previous discussion of the stepwise extension of sections and look at the first dimension in which $\pi_{k-1}(F_x) \neq 0$. If $s_1$ and $s_2$ are two extensions from sections on the $k$-1-skeleton to sections on the $k$-skeleton then one can define a cocycle:

$$d(s_1, s_2) \in H^{k-1}(M; \pi_{k-1}(F_x))$$

that one calls the *primary difference cocycle.* It has the property that if $\mathfrak{o}_1$ and $\mathfrak{o}_2$ are the obstruction cocycles for each section at this stage then:

---

[1] For the sake of completeness, a brief Appendix on the subject of CW-complexes is included at the end of the article.

$$\delta d(s_1, s_2) = o_1 - o_2.$$

The sections are homotopic iff $d(s_1, s_2) = 0$. As a result, one can parameterize the homotopy classes of sections by the elements of $H^{k-1}(M; \pi_{k-1}(F_x))$; in particular, if this module vanishes then all such sections are homotopic.

One of the early applications of this process − or, perhaps, an early application that later *led* to this process − was Stiefel's work [**18**] on the parallelizability of differentiable manifolds. The form that obstruction theory takes in this context is that of looking for obstructions to the existence of global fields of $k$-frames, i.e., the existence of $k$ nonzero vector fields $\mathbf{e}_i$, $i = 1, \dots k$ on the manifold M that are linearly independent at every point. Note that this is a recursive problem, in the sense that first one must find a global non-zero vector field, then a global 2-frame field, etc. Hence, the primary obstruction to the existence of a global $k$-frame field must be added to the set of obstructions that had to vanish in order to define $m$-frame fields for all $m < k$.

In order to define the fiber bundle that puts the aforementioned into the form of obstruction theory, one first defines the $k^{\text{th}}$ *Stiefel manifold* $V_{n,k}$, which consists of all orthonormal ($^2$) $k$-frames in an $n$-dimensional Euclidean vector space. This set is given the topology and differential structure that it obtains when one expresses it as the homogeneous space $O(n)/O(n-k)$. (This amounts to the statement that any $n$-frame can be split into a $k$-frame and an $n-k$-frame, but the $n-k$-frame part is irrelevant when one considers only the $k$-frame.) The first non-trivial homotopy group of $V_{n,k}$ is:

$$\pi_{n-k}(V_{n,k}) = \begin{cases} \mathbb{Z} & k = 1 \text{ or } n-k \text{ even} \\ \mathbb{Z}_2 & k \geq 2 \text{ and } n-k \text{ odd}. \end{cases}$$

As an example of the application of obstruction theory, we shall follow the construction of a 4-frame field on a four-dimensional M − such as spacetime, presumably − and look at the individual obstructions:

1-frame field, i.e., nonzero vector field:    $V_{4,1} = O(4)/O(3) = S^3$, so the first non-trivial homotopy group is $\pi_3(V_{4,1}) = \mathbb{Z}$, and the obstruction is an element of $H^4(M; \mathbb{Z})$. When M is orientable and oriented this 4-cocycle is the Euler class of M, and its $\mathbb{Z}_2$-reduction is called the fourth *Stiefel-Whitney class* of $T(M)$, which is denoted by $w_4$. The homotopy classes of non-zero vector fields are then parameterized by the elements of $H^3(M; \mathbb{Z})$. In the compact oriented case, Poincaré duality makes this isomorphic to $H_1(M; \mathbb{Z})$ and the Hurewicz isomorphism makes this isomorphic to the Abelianization of $\pi_1(M)$. Hence, a simply connected manifold will have one homotopy class of nonzero vector field on it.

2-frame field: $V_{4,2} = O(4)/O(2)$, whose first non-trivial homotopy group is $\pi_2(V_{4,2}) = \mathbb{Z}$, so the obstruction is an element $w_3 \in H^3(M; \mathbb{Z})$ whose $\mathbb{Z}_2$-reduction is called the third

---

$^2$ Although the case of *linear $k$-frames* would seem more general, in the eyes of homotopy groups, the essential information is carried by the maximal compact subgroup of $GL(n)$, namely $O(n)$.

Stiefel-Whitney class. The homotopy classes of 2-frame fields are then parameterized by the elements of $H^2(M; \mathbb{Z})$.

3-frame field: $V_{4,3} = O(4)/O(1) = \mathbb{R}P^3$, whose first non-trivial homotopy group is $\pi_1(V_{4,3}) = \mathbb{Z}_2$, and this gives us the second Stiefel-Whitney class $w_2 \in H^2(M; \mathbb{Z}_2)$. The homotopy classes of 3-frame fields are then parameterized by the elements of $H^1(M; \mathbb{Z}_2)$. This means that if M is simply connected or if $\pi_1(M)$ has no elements of order 2 then all 3-frame fields on M are homotopic.

4-frame field: $V_{4,4} = O(4)/O(0) = O(4)$, whose first non-trivial homotopy *set* is $\pi_0(V_{4,4}) = \mathbb{Z}_2$, so the final obstruction is the first Stiefel-Whitney $w_1 \in H^1(M; \mathbb{Z}_2)$. The homotopy classes of 4-frame fields are then parameterized by the elements of $H^0(M; \mathbb{Z}_2)$. A path-connected M will have two such classes, corresponding to left and right orientations for the frames.

The first Stiefel-Whitney class is also involved with the orientability of T(M); in fact, for a compact M, T(M) is orientable iff $w_1 = 0$ (cf. Husemoller [**15**]).

Stiefel showed that any compact orientable 3-manifold is completely parallelizable. Since the vanishing of $w_2$ is also necessary and sufficient for the existence of an $SL(2; \mathbb{C})$-spin structure on M, it is not surprising that Geroch [**19**] was able to show that if M is a non-compact orientable four-dimensional Lorentz manifold then it admits such a spin structure iff it is parallelizable. One also has that the homotopy classes of spin structures are parameterized by the elements of $H^1(M; \mathbb{Z}_2)$. (We have already seen that the orientation requires that $w_1 = 0$ and the Lorentz structure implies that $w_4 = 0$. However, non-compactness would make $H^4(M) = 0$ anyway.)

Since complete parallelizability implies a lot of symmetry – indeed, one might call completely parallelizable manifolds "almost Lie groups" – this suggests an interesting dilemma to resolve in physics, given the importance of spinors in quantum field theory. Either one must regard the $SL(2; \mathbb{C})$-principal bundle as the fundamental object and consider its (singular) projection onto the bundle of linear frames as the classical approximation, or turn to the use of *generalized spin structures* [**20**].

For the case of H-reductions of G-principal bundles the fiber is G/H, so the obstructions will be strongly influenced by the homotopy groups of this homogeneous space, as well as the cohomology of M. For instance, when one reduces from GL(M) to O(M), i.e., defines a Riemannian metric on M, the homogeneous space in question is GL(4)/O(4), which is diffeomorphic to $\mathbb{R}^{10}$. Since this is a contractible space, there is no dimension in which there is a non-vanishing obstruction cocycle for this reduction. This is simply a fancy way of restating the fact that any paracompact differentiable admits a Riemannian metric. Since paracompactness played no role in the foregoing discussion, one must remember that the vanishing of the obstruction cocycle is generally necessary, but not sufficient, for the reduction to take place; paracompactness is a sufficient condition, as well as a necessary one.

The fact that the homogeneous space GL(4)/O(4) is contractible is actually due to a more elementary topological fact: O(4) is a deformation retract of GL(M), where a subset A in a topological space X is called a *deformation retract* of X iff there is a surjection $p$:X → A, called a *deformation retraction,* such that its composition with the inclusion $p·i$ is homotopic to the identity. A useful property of any deformation retraction is that it preserves all homotopy groups. Hence, if H is a deformation retract of G then $\pi_i$(G) ≅ $\pi_i$(H) for all $i$, and a straightforward application of the homotopy exact sequence for H → G → G/H will give the triviality of all $\pi_i$(G/H), which gives the contractibility of G/H ([3]).

The situation for a four-dimensional Lorentz manifold is slightly more involved, since the homogeneous space GL(4)/O(3,1), which represents the space of Lorentz scalar products on Minkowski space, is diffeomorphic to ($\mathbb{R}^{10} \times$O(4))/($\mathbb{R}^3 \times$O(3)), which is homotopically equivalent to $\mathbb{R}P^3$. Hence, defining a Lorentz metric on M is homotopically equivalent to defining a global line field on M. The *primary* obstruction to this is an element of H$^2$(M; $\pi_1(\mathbb{R}P^3)$), which nevertheless vanishes (cf. Steenrod [16]), so we go on to the *secondary* obstruction in H$^4$(M; $\pi_3(\mathbb{R}P^3)$) = H$^4$(M; $\mathbb{Z}$). Since this module vanishes if M is non-compact, any non-compact differentiable manifold admits a Lorentz structure, but when M is compact the vanishing of the obstruction cocycle obstruction is equivalent to the vanishing of the Euler-Poincaré characteristic. The homotopy classes of Lorentz structures on M are then related to H$^3$(M; $\mathbb{Z}$); again, for the oriented case, this goes back to the abelianization of $\pi_1$(M), and a simply connected Lorentz manifold will admit one homotopy class of Lorentz structure.

Notice that this last example shows that a reduction does not have to be associated with a contractible homogeneous space – i.e., a vector space, up to homotopy − in order for it to have a fundamental tensor field.

One should point out that since a reduction to a G-structure also can be considered as a choice of G-orbit in GL(M) at each point, and there is nothing unique about these G-orbits in each fiber of GL(M), then there is also nothing unique about a given G-structure. In particular, every frame in a given fiber GL$_x$(M) generates a G-orbit. For instance, any linear frame $\mathbf{e}_i \in$GL$_x$(M) can be considered to be an orthonormal frame for some Riemannian metric:

$$g = \theta^i \otimes \theta^i,$$

where $\theta^i$ is the reciprocal coframe field to $\mathbf{e}_i$:

$$\theta^i(\mathbf{e}_j) = \delta^i_j.$$

---

[3] The fact that the vanishing of all homotopy groups implies the contractibility of a topological space is less straightforward than it sounds - as opposed to its converse - and is referred to as Whitehead's Theorem [29, 30]

This metric is also characterized as the one that makes the frame $\mathbf{e}_i$ orthogonal. Note that any other frame in $GL_x(M)$ that is related to $\mathbf{e}_i$ by an orthogonal transformation will define the same metric at $x$. This why the use of "vierbeins" to define a spacetime metric should include the cautionary note that the metric is associated with an *equivalence class* of frames *up to isometry*, since a manifold may not actually admit a global frame field, even when it does admit a global metric.

**2. Minkowski space preliminaries.** Before we discuss reductions of the bundle of affine frames on spacetime, we will examine a particular chain of subgroups in A(4), the affine group of $\mathbb{R}^4$, that are associated with giving $\mathbb{R}^4$ the structure of Minkowski space.

The chain that we shall consider is:

$$A(4) \leftarrow GL(4) \leftarrow GL^+(4) \leftarrow SL(4) \leftarrow SO(3,1) \leftarrow SO_0(3,1) \leftarrow SO(3) \leftarrow SO(2) \leftarrow \mathbb{Z}_2 \leftarrow \{e\},$$

in which the arrows represent subgroup inclusions. Naturally, the choice of representative for each subgroup is not always unique, but we shall not require that to be the case, anyway.

First we shall go through this sequence step-by-step and discuss the homogeneous space that one obtains, as well as the interpretation of the step in terms of frames in Minkowski space. Then we shall apply the results to the case of the bundle of affine frames on a connected four-dimensional manifold without boundary.

$A(4) \leftarrow GL(4)$: The homogeneous space $A(4)/GL(4)$ is diffeomorphic to $\mathbb{R}^4$, since each element of it represents a choice of origin in $A^4$. Because $\mathbb{R}^4$ is a contractible space, we have:

$$\pi_i(A(4)/GL(4)) = 0, \text{ for all } i.$$

Since A(4) is a semi-direct product of the groups $\mathbb{R}^4$ and GL(4), the underlying manifold is simply the product manifold $\mathbb{R}^4 \times GL(4)$ and the deformation retraction in this case is the projection onto GL(4). In terms of frames, we have simply replaced an affine frame $\mathbf{a}_i$ with a pair $(\mathbf{o}, \mathbf{e}_i)$, where $\mathbf{o} \in A^4$ and $\mathbf{e}_i$ is a frame in $\mathbb{R}^4$.

$GL(4) \leftarrow GL^+(4)$: The homogeneous space $GL(4)/GL^+(4)$ is diffeomorphic to $\mathbb{Z}_2$, which we represent as the set $\{-1, +1\}$, whose elements represent the possible choices of sign for det A, when $A \in GL(4)$. Such a choice of sign amounts to a choice of orientation for a frame, which one sometimes refers to as "left-handed" and "right-handed," respectively. The only difference between the homotopy groups of $\mathbb{Z}_2$ and the homotopy groups of a point is in the fact that $\mathbb{Z}_2$ has two path-connected components; hence:

$$\pi_0(GL(4)/GL^+(4)) = \mathbb{Z}_2$$
$$\pi_i(GL(4)/GL^+(4)) = 0, \text{ for all } i > 0.$$

Another way of looking at this step is to see that GL(4) has two connected components: $GL^+(4)$, which is a subgroup that contains the identity element, and a diffeomorphic copy that can be obtained by multiplying each element of $GL^+(4)$ by any matrix in GL($n$) that has a negative determinant. Some particularly interesting cases, in the eyes of physics are the matrices that reflect any odd combination of frame members through the origin. For instance, if the $0^{th}$ frame member represents the generator of the proper time axis then one such choice is the matrix of *time inversion:*

$$T = \begin{bmatrix} -1 & 0 & 0 & 0 \\ 0 & 1 & 0 & 0 \\ 0 & 0 & 1 & 0 \\ 0 & 0 & 0 & 1 \end{bmatrix} = \text{diag}[-1, 1, 1, 1];$$

another choice is *spatial inversion:*

$$P = \text{diag}[1, -1, -1, -1].$$

Note that $PT = TP = -I$, i.e., complete reflection through the origin, is again an element of $GL^+(4)$. This is distinct from the analogous situation in three dimensions, where a reflection through the origin would invert the orientation of a frame. Collectively, the four elements $\{I, T, P, TP = -I\}$ define a group that is (isomorphic to) the *Klein viergruppe* $\mathfrak{V}$. It is commutative, non-cyclic, and all of its elements are *involutory;* i.e., $a^2 = I$, for all $a \in \mathfrak{V}$.

$GL^+(4) \leftarrow SL(4)$: $GL^+(4)/SL(4)$ is diffeomorphic to $\mathbb{R}^+$, since all we are doing is factoring out the determinant of any matrix: $A = |A| A_1$, where $A_1 \in SL(4)$. In terms of frames, what we have done is to choose a scale of unit *volume* for a frame. It is important to notice that this is still *in advance of any choice of metric.* Since $\mathbb{R}^+$ is contractible the homotopy groups all vanish.

$$\pi_i(GL^+(4)/SL(4)) = 0, \text{ for all } i.$$

This reduction is a deformation retraction that is essentially the projection $GL^+(n) = \mathbb{R}^+ \times SL(n) \rightarrow SL(n)$ that is given by $A \mapsto \dfrac{1}{|A|} A$.

$SL(4) \leftarrow SO(3,1)$: $SL(4)/SO(3,1)$ is best represented by volume-preserving *Lorentz strains*. They are invertible 4×4 real matrices with the property that they are *Lorentz self-adjoint*:

$$S^* = S,$$

where:

$$S^* \equiv \eta S^T \eta,$$

in which $\eta$ is the matrix of the Lorentz scalar product.

One can decompose any element $A \in GL(4)$ into the product of a Lorentz self-adjoint matrix and a Lorentz transformation:

$$A = \Sigma L$$

in a manner that is analogous to the decomposition of such a matrix into the product of a positive-definite symmetric matrix and an orthogonal matrix. (The proof is essentially by Wick conjugation of the orthogonal case.) In terms of frames, it is better to think of the elements of $\Sigma_0 = SL(4)/SO(3,1)$ as volume-preserving *deformations* of the scalar product $\eta$, with the action given by:

$$\eta \mapsto \Sigma^T \eta \, \Sigma,$$

which is also the scalar product that one obtains when one deforms a frame $\mathbf{e}_i$ in $\mathbb{R}^4$ by $\mathbf{e}_i \mapsto A\mathbf{e}_i$, and decomposes A into $\Sigma L$.

The elements of $\Sigma_0$ correspond to (isometrically) distinct choices of Lorentz scalar product for Minkowski space. In terms of frames, since any frame in $\mathbb{R}^4$ spawns an $SO(3,1)$-orbit, a choice of scalar product corresponds to a choice of such an orbit. The homotopy groups of $\Sigma_0$ are obtained by noting that the underlying manifold of $SL(4)$ is $\mathbb{R}^9 \times SO(4)$ and the underlying manifold of $SO(3,1)$ is $\mathbb{Z}_2 \times \mathbb{R}^3 \times SO(3)$ so the manifold $\Sigma_0$ is diffeomorphic to $\mathbb{R}^6 \times SO(4)/\mathbb{Z}_2 \times SO(3) = \mathbb{R}^6 \times \mathbb{R}P^3$:

$$\pi_i(SL(4)/SO(3,1)) = \pi_i(\mathbb{R}P^3) = \begin{cases} 0 & i = 0, 2 \\ \mathbb{Z}_2 & i = 1 \\ \mathbb{Z} & i = 3, 4 \end{cases}.$$

This is what accounts for the fact that choosing a Lorentz scalar product is equivalent (up to homotopy) to choosing a line through the origin of $M^4$. Since the physical interpretation of this choice is that it represents a choice of proper-time axis, i.e., a choice of rest frame ([4]), it is easy to see that this choice is not actually canonical. Hence, in the geometrical context, one still must choose both the scalar product and the proper-time axis independently. Geometrically, one must still account for the $\mathbb{R}^6$ factor in $\Sigma_0$, which represents the Lorentz strains that act on the choice of proper time axis.

$SO(3,1) \leftarrow SO_0(3,1)$: The Lorentz group $O(3,1)$ is not connected, but consists of four diffeomorphic components, which correspond to the four transformations I, T, P, and TP $= -I$. Once we have fixed an orientation for $M^4$, we have reduced this to the two components of $SO(3,1)$: the connected component of the identity $SO_0(3,1)$, which is usually called the *proper orthochronous Lorentz group* – or also the *restricted Lorentz group* – and its image under the TP transformation. This, in turn, corresponds to the two choices of unit vector in the proper-time axis, i.e., to a choice of *time orientation* (relative to a choice of spatial orientation); this also accounts for the factor of $\mathbb{Z}_2$ in the decomposition of $SO(3,1)$ as a product manifold. Here, we are dealing with the

---

[4] A *rest frame* is actually an *equivalence class* of orthonormal frames that share a common timelike member, under the action of $SO(3)$ in the spacelike subspace that is orthogonal to that timelike vector.

homogeneous space $SO(3,1)/SO_0(3,1)$, which is $\mathbb{Z}_2$, up to diffeomorphism. Hence, we have:

$$\pi_0(SO(3,1)/SO_0(3,1)) = \mathbb{Z}_2$$
$$\pi_i(SO(3,1)/SO_0(3,1)) = 0, \text{ for all } i > 0.$$

$SO_0(3,1) \leftarrow SO(3)$: $SO_0(3,1)/SO(3)$ is the set of all boosts of a chosen rest frame. Since this also corresponds to a choice of proper-time axis, one sees that, in effect, this reduction always follows the previous one. By the usual decomposition (i.e., the Gram-Schmidt one in $GL(4)$) of an element of $SO(3,1)$ into a positive definite symmetric matrix (with determinant one) and an element of $SO(3)$, one sees that $SO(3,1)/SO(3)$ is diffeomorphic to $\mathbb{R}^3$. The reduction of frames that we have just made is from the set of all Lorentz-orthonormal 4-frames that were chosen in the previous step to the corresponding set of Euclidean orthonormal spacelike 3-frames that are orthogonal to the proper-time axis. Once again, we have a contractible homogeneous space, so:

$$\pi_i(SO(3,1)/SO(3)) = 0, \qquad \text{for all } i.$$

The deformation retraction in this case is another projection, once one has factored an element of $SO_0(3,1)$ into the product of a boost and a rotation.

$SO(3) \leftarrow SO(2)$: The homogeneous space $SO(3)/SO(2)$ is diffeomorphic to $S^2$. Hence, an element of this space can be regarded as a choice of spacelike unit vector in the aforementioned spacelike Euclidean 3-plane in $M^4$. This not only reduces the previous set of spacelike orthonormal 3-frames to a set of spacelike orthonormal 2-frames, but since we have already fixed a timelike unit vector in order to define a time orientation and the spacelike unit vector at this step, we have also defined a timelike orthonormal 2-frame. The homotopy groups in question are:

$$\pi_i(SO(3)/SO(2)) = \pi_i(S^2) = \begin{cases} 0 & i = 0,1 \\ \mathbb{Z} & i = 2,3 \end{cases}.$$

$SO(2) \leftarrow \mathbb{Z}_2$: The $\mathbb{Z}_2$ subgroup of $SO(2)$ consists of $\{I, R_\pi\}$, namely the identity rotation and a rotation by $\pi$ radians. These two points span a line through the origin in the plane, and $SO(2)/\mathbb{Z}_2$ then amounts to all lines through the origin of the spacelike 2-plane that we obtained in the previous step, i.e., $SO(2)/\mathbb{Z}_2$ is diffeomorphic to $\mathbb{R}P^2$. In terms of frames, we are reducing the set of all spacelike orthonormal 2-planes that we obtained from the previous step to a pair of spacelike unit vectors in the chosen line through the origin. The homotopy groups for this case are:

$$\pi_i(SO(2)/\mathbb{Z}_2) = \pi_i(\mathbb{R}P^2) = \begin{cases} 0 & i \neq 1 \\ \mathbb{Z}_2 & i = 1 \end{cases}.$$

$\mathbb{Z}_2 \leftarrow I$: This otherwise trivial step amounts to making the actual choice of unit vector in the spacelike line that was just chosen. Note that so far we have explicitly chosen three orthonormal vectors up to this point in the chain, whereas a reduction of $GL(4)$ to $\{e\}$ amounts to a choice of orthonormal 4-frame. The reason that there is no contradiction in saying this is that choosing three orthonormal vectors defines a line that is orthonormal to the 3-plane that they span, and since we have already chosen an orientation for $M^4$, an orientation for the proper-time axis, and an orientation for the chosen spacelike unit vector, the choice of orientation for the remaining spacelike unit vector is unique.  The only non-trivial homotopy group here is $\pi_0(\mathbb{Z}_2)$.

Some might suggest that the chain of subgroups that we have chosen omits the widely discussed case of the linear conformal group for the Lorentz group CO(3,1), or the linear conformal groups that are associated with the lower-dimensional metric structures, SO(3) and SO(2).  Certainly we are not suggesting that the chosen chain of subgroups is the only one possible.  For instance, the sequence of reductions $GL^+(4) \leftarrow SL(4) \leftarrow SO(3,1)$ could have gone through the sequence $GL^+(4) \leftarrow CO(3,1) \leftarrow SO(3,1)$.

For the record, if (V, $g$) is an orthogonal space, in which the scalar product $g$ can have a general signature, then a *conformal transformation* of (V, $g$) is a diffeomorphism A of V such that there is a $\lambda^2 > 0$ such that for all $\mathbf{v}$, $\mathbf{w} \in$ V:

$$g(A\mathbf{v}, A\mathbf{w}) = \lambda^2 g(A\mathbf{v}, A\mathbf{w}).$$

As opposed to rigid motions of (V, $g$), which are all affine transformations, the conformal transformations of (V, $g$) can include nonlinear diffeomorphisms as well as affine ones.  For instance, the conformal group of Minkowski space is composed of not only the Poincaré group and the dilatations but also four nonlinear inversions:

$$\mathbf{v} \mapsto -\frac{1}{\| \mathbf{v} \|}\mathbf{v},$$

that can also represent transformations to frames that have a constant relative acceleration.  A popular source of research directions is the nonlinear conformal group for a two-dimensional Euclidean space, which is an *infinite*-dimensional group; indeed, two is the only dimension in which the conformal group is not finite-dimensional.

Although one can often find faithful linear representations of the nonlinear conformal group in higher dimensions, we shall refer to the subgroup of the conformal group for (V, $g$) that lies in GL(V) as the *linear conformal group*.  If the orthogonal group for $g$ is O($k$, $n$-$k$) then the linear conformal group that goes with CO($k$, $n$-$k$) it is isomorphic to $\mathbb{R}^+ \times$O($k$, $n$-$k$), so a typical element A$\in$CO($k$, $n$-$k$) can also be described by the pair ($\lambda$, R), where $\lambda = |$det A$|$ and R $= \lambda^{-1}$A.  A *conformal frame* in V is an orthogonal frame for the scalar produce $g$ whose elements all have equal length.  If one such frame is chosen to have unit length members then all of the other conformal frames are obtained from it by either orthogonal transformations or dilatations.  Hence, one could regard the space of conformal frames an extension of any choice of orbit for the orthogonal group by the action of the group $\mathbb{R}^+$.

We have encountered the same factor $\lambda$ in the decomposition an element of $GL^+(4)$ into the product of a dilatation, Lorentz shear, and Lorentz transformation.  Hence, one can see that since there are two projections onto the Lorentz group there are two directions of symmetry breaking. However, since both the dilatations and the Lorentz shears are part of the group $GL^+(4)$ that does not affect the homotopy type of the reduced subgroup, the approach that we shall take in what follows is to consider the essential links in the chain to be the subgroups that are homotopically distinct. We can then regard the other subgroups as deformations of the essential ones in a manner analogous to the way that one might consider the states within a given phase of condensed matter as excitations of a ground state in that phase.

For instance, the orthogonal subgroups in our chain that are susceptible to deformation by the dilatation group are $SO(3,1)$, $SO_0(3,1)$, $SO(3)$, and $SO(2)$. In the last case, one should point out that the resulting conformal group is isomorphic to $\mathbb{C}^*$, the multiplicative group of non-zero complex numbers, i.e.:

$$CO(2) \cong \mathbb{R}^+ \times SO(2) \cong \mathbb{C}^*, \qquad A \mapsto (r, \theta) \mapsto re^{i\theta}.$$

We note in passing that a similar observation pertains to the case of deforming $S^3$ − in the form of $SU(2)$ and represented by unit quaternions − by dilatations, to obtain the multiplicative quaternion group:

$$\mathbb{R}^+ \times SU(2) \cong \mathbb{Q}^*.$$

In a previous work of the author [8], the role of deformations of the Minkowski scalar product by $\mathbb{C}^*$ was seen to account for the Madelung potential in the real form of the Schrödinger equation as an artifact of the appearance of non-zero curvature in the conformally-deformed Levi-Civita connection. Some of the motivation for the present work was to put this result into a broader mathematical context and explore the surrounding ideas for more such insight.

To summarize: the chain of subgroups:

$A(4) \leftarrow GL(4) \leftarrow GL^+(4) \leftarrow SL(4) \leftarrow SO(3,1) \leftarrow SO_0(3,1) \leftarrow SO(3) \leftarrow SO(2) \leftarrow \mathbb{Z}_2 \leftarrow \{e\},$

gives rise to the sequence of homogeneous spaces:

$\mathbb{R}^4, \quad \mathbb{Z}_2, \quad \mathbb{R}^+, \quad \mathbb{R}^6 \times \mathbb{R}P^3, \quad \mathbb{Z}_2, \quad \mathbb{R}^3, \quad S^2, \quad \mathbb{R}P^2, \quad \mathbb{Z}_2,$

which one geometrically interprets as:

A choice of origin in $A^4$,
A choice of left/right orientation for the frames in $\mathbb{R}^4$,
A choice of unit volume for the oriented frames in $\mathbb{R}^4$,
A choice of Lorentz scalar product and proper time axis for $\mathbb{R}^4$,
A choice of orientation for the proper time axis,
A choice of vector in the proper time axis,
A choice of unit vector in the proper time axis,
A choice of spacelike line in the subspace orthogonal to the proper time line,
A choice of orientation for the spacelike line (hence, the choice of unit vector that generates it) .

The topologically essential subchain of the above – i.e., the chain of subgroups in which the homotopy type of the homogeneous space is non-trivial is:

$$GL(4) \leftarrow GL^{+}(4) \leftarrow SO(3,1) \leftarrow SO_0(3,1) \leftarrow SO(2) \leftarrow \mathbb{Z}_2 \leftarrow \{e\},$$

with corresponding homogeneous spaces (up to homotopy):

$$\mathbb{Z}_2, \quad \mathbb{R}P^3, \quad \mathbb{Z}_2, \quad S^2, \quad \mathbb{R}P^2, \quad \mathbb{Z}_2.$$

We shall regard the *phases* of the spacetime vacuum manifold as any of these homogeneous spaces. The intermediary reductions, whose homogeneous spaces are vector spaces, up to homotopy, represent deformations or excitations of the vacuum "ground state" in the phase. In other words, they are changes of state that do not change the phase, somewhat like raising the temperature of a solid without melting it.

Let us pause to clarify the term *ground state* in the context of the phases of the spacetime vacuum manifold. Since the point we made above was that reductions of symmetry groups were associated with two types of homogeneous spaces − contractible and non-contractible ones – and contractions have to terminate in constant maps, which correspond to deformation retracts of the G-structures, we are saying that the ground state in a given phase G(M) of the spacetime vacuum manifold is any reduction H(M) of G(M) to an H-structure with G/H non-contractible. This ground state does not have to be unique though, since homotopically distinct H-reductions of a G-structure are possible. Hence, the ground state of a given phase of the spacetime vacuum manifold can be degenerate.

If we follow the sequence of reductions under present consideration, we see that the deformations take the form of translations of linear frames within the affine frame bundle, dilatations, Lorentz shears, and boosts of rest frames. Any of these deformations can be applied to the phases of the spacetime vacuum manifold, in the context of 4×4 real matrices, but only some deformations would seem to have any immediate physical significance, such as dilatations deforming any of the subgroups below $GL^{+}(4)$; of course, that could just be a first impression.

**3. Spacetime G-structures.** Now, we shall go over the same chain of subgroups and homogeneous spaces that we discussed in the last section, but in the context of the bundle of affine frames on spacetime this time, and discuss the mathematical nature of the topological obstructions to the reductions at each step and the fundamental tensor fields that are associated with the reduction – if they exist. The reason we shall still include the topologically trivial reductions is because they sometimes still produce useful geometric information, such as fundamental tensor fields.

$A(M) \leftarrow GL(M)$: Since $G/H = \mathbb{R}^4$ is contractible in this case, there is no topological obstruction to this step. The fundamental tensor field $t$: $A(M) \rightarrow \mathbb{R}^4$ amounts to the "soldering form" one associated with a Cartan connection, which becomes the canonical 1-form $\theta^i$ on GL(M) after reduction. The fact that there is no obstruction to this reduction is consistent with the conventional wisdom that says an affine connection on a manifold M can be replaced by a linear connection that has the same geodesics.

GL(M) ← GL$^+$(M): This time, G/H $= \mathbb{Z}_2$, which has non-trivial homotopy only in dimension zero. Hence, the obstruction to the reduction is the orientability of T(M), which takes the form of a 1-cocycle with values in $\pi_0(\mathbb{Z}_2)$:

$$\mathit{o}_1(M) \in H^1(M, \mathbb{Z}_2);$$

in fact, this obstruction is the first Stiefel-Whitney class of M, $w_1$. As mentioned before, for a compact M, $w_1 = 0$ iff T(M) is orientable. In particular, a simply connected M will be orientable. (One could also derive this latter fact from a universal orientable covering manifold argument.) Furthermore, since the homotopy classes of orientations are parameterized by $H^0(M, \mathbb{Z}_2)$, if M is path-connected then there will be two homotopy classes of orientations.

GL$^+$(M) ← SL(M): Here, G/H is diffeomorphic to $\mathbb{R}^+$, which is contractible, so there is no topological obstruction to the reduction. We shall denote the fundamental tensor field by det: GL$^+$(M) $\to \mathbb{R}^+$. It may be regarded as either a smooth function on GL$^+$(M) that is constant on the orbits of SL(4) or a non-zero 4-form V on M. The representation of det for each frame $\mathbf{e}_i \in GL^+_x(M)$ is equal to the determinant of the matrix of that frame with respect to a fixed, but arbitrary, reference frame in $T_x(M)$, an expression that is well-defined even when M is not parallelizable, since the determinant of that matrix is independent of choice of reference frame. If $\theta^i$ is the coframe that is reciprocal to $\mathbf{e}_i$ then the map:

$$V: GL^+(M) \to \Lambda^4(\mathbb{R}^4), \qquad \mathbf{e}_i \mapsto \det(\mathbf{e}_i)\theta^0 \wedge \theta^1 \wedge \theta^2 \wedge \theta^3,$$

is well-defined at each $x \in M$, i.e., it is independent of the choice of frame. Since det is constant on each SL(4)-orbit, once we have chosen an SL(4)-orbit at each $x \in M$ we have implicitly defined a smooth function on M, which we denote by $\rho = \det(\mathbf{e}_i)$, for any $\mathbf{e}_i$ in the SL(4) orbit at $x$.

SL(M) ← SO(3,1)(M): For this step, G/H $= \mathbb{R}^6 \times \mathbb{RP}^3$, which is homotopically equivalent to $\mathbb{RP}^3$. As discussed above, defining a Lorentz metric on a manifold is topologically equivalent to defining a global section of the projectivized tangent bundle on M, i.e., a global line field. As mentioned before, the primary obstruction is an element of $H^2(M; \pi_1(\mathbb{RP}^3)) = H^2(M; \mathbb{Z}_2)$, but this actually vanishes, so we go to the secondary obstruction in $H^4(M; \pi_3(\mathbb{RP}^3)) = H^4(M; \mathbb{Z})$. Since we have already chosen an orientation for T(M), this is the Euler class of T(M), $e[M]$. For a non-compact M, $H^4(M) = 0$ and there is no obstruction, but for a compact M, the obstruction is determined by the Euler-Poincaré characteristic $\chi(M)$.

The homotopy classes of Lorentz structures on a compact M are indexed by the elements of $H^3(M; \mathbb{Z})$. As mentioned before, since we have already defined an orientation to get this far, we can use Poincaré duality to associate $H^3(M; \mathbb{Z})$ with $H_1(M; \mathbb{Z})$, which is isomorphic to $\pi_1(M)$ by the Hurewicz isomorphism. Hence, a simply connected M will have one homotopy class of Lorentz structure.

The fundamental tensor field for this reduction is, of course, a Lorentz metric on M:

$$g: SL(M) \rightarrow \Sigma_0.$$

$g$ also defines a bundle isomorphism of GL(M) with GL*(M) by means of ([5]):

$$\mathbf{e}_i \mapsto e^i = i_{\mathbf{e}_i} g = g_{ij} \theta^j.$$

Hence, this isomorphism is generally different from the one that associates the reciprocal coframe to a frame, but for $g$-orthonormal frames, they will differ only by an orientation. Note that under this isomorphism, when it is extended to the exterior algebras over T(M) and T*(M), the 4-form $\mathbf{e}_0 \wedge \mathbf{e}_1 \wedge \mathbf{e}_2 \wedge \mathbf{e}_3$ goes to:

$$e^0 \wedge e^1 \wedge e^2 \wedge e^3 = (g_{0j} \theta^j) \wedge (g_{1j} \theta^j) \wedge (g_{2k} \theta^k) \wedge (g_{3l} \theta^l) = \frac{1}{4!} \varepsilon_{ijkl} \, (g_{0i} \, g_{0j} \, g_{0k} \, g_{0l}) \, \theta^0 \wedge \theta^1 \wedge \theta^2 \wedge \theta^3$$

$$= \det(g) \, \theta^0 \wedge \theta^1 \wedge \theta^2 \wedge \theta^3.$$

Hence, we can use the metric tensor field as a way of defining the volume element. We shall see that as we go down the chain of subgroups, we can generally define all of the geometrical information above a given level of that hierarchy using the information that was introduced by the previous reductions.

$SO(3,1)(M) \leftarrow SO_0(3,1)(M)$: Since G/H is diffeomorphic to $\mathbb{Z}_2$, we are looking at an obstruction to the orientability of a bundle. The difference is that we are not orienting T(M) this time, but the *line* bundle that is defined by a choice of line field L(M) on M, and it is entirely possible for orientable manifolds to admit non-orientable line fields. Hence, one should be careful to distinguish between $w_1$ for T(M) and $w_1$ for L(M), which we denote by $w_1[L]$. One generally refers to a choice of orientation on L(M) as a (proper) *time orientation* on M. Again, any simply connected M will also be time orientable; in fact, one can also construct a time orientable universal covering manifold for a Lorentz manifold M. Analogous to what we saw for the orientations of T(M), if M is path-connected then there will be two homotopy classes of time orientations.

$SO_0(3,1)(M) \leftarrow SO(3)(M)$: This time, G/H is diffeomorphic to $\mathbb{R}^3$, which is contractible, so there is no topological obstruction to the reduction. The fundamental tensor field that is defined by this reduction is a timelike non-zero vector field $\mathbf{t}$ on M, which is also sometimes used as the definition of a time orientation. Actually, to be precise, since we are looking at equivalence classes of real 4×4 matrices, we should briefly pause to show how one gets from an equivariant map on $SO_0(3,1)(M)$ with values in such an equivalence class to a vector field on M. Mostly, one needs to know that although the tangent spaces to G/H do not define a Lie subalgebra of $SO_0(3,1)$, nevertheless, they define a three-dimensional vector subspace of that Lie algebra, namely the space of infinitesimal boosts along spacelike directions. Moreover, G/H is parallelizable in this particular case, so the tangent vectors to G/H, in particular, the values of the fundamental tensor field $t$, can be expressed in the form $t^i L_i$, $i = 1, 2, 3$, where the $L_i$ represent the matrices of the elementary infinitesimal boosts along the $x, y,$ and $z$-direction in Minkowski space, relative to the canonical frame on $\mathbb{R}^4$. Now, if $\mathbf{e}_j, j = 0, 1, 2, 3$, is an element of $SO_0(3,1)(M)$ in the fiber at $x{\in}M$ and $\mathbf{e}_0$ is the future-pointing timelike member of the frame then we define the vector field $\mathbf{t}$ at each $x{\in}M$ by:

$$\mathbf{t} = \mathbf{e}_0 + t^i \mathbf{e}_i, \qquad i = 1, 2, 3.$$

---

[5] Although the placement of the indices looks inconsistent, this is one of the rare times that it is not.

(This is where the equivariance of the map $t$ is necessary; otherwise, a different choice of $\mathbf{e}_i$ at $x{\in}M$ would give a different vector in $T_x(M)$.) Note that the rest frame for this reduction consists of all the frames that map to 0 under $t$.

Sometimes, instead of looking at the vector field $\mathbf{t}$, we shall also consider the equivariant 0-form $t^i : SO_0(3,1)(M) \to \mathbb{R}^3$.

Now that we have the unit timelike vector field $\mathbf{t}$, in addition to V and $g$, we can define a number of other geometric objects:

| | |
|---|---|
| A unit timelike 1-form: | $\theta^0 = i_{\mathbf{t}} g$ |
| A Riemannian metric on $\Sigma(M)$: | $g_\Sigma = \theta^0 {\otimes} \theta^0 - g$ |
| A volume element on $\Sigma(M)$: | $V_\Sigma = i_{\mathbf{t}} V.$ |

We can go the other direction in the chain by starting with $\theta^0$, $g_\Sigma$, and $V_\Sigma$, and defining:

$g = \theta^0 {\otimes} \theta^0 - g_\Sigma,$
$V = \theta^0 {\wedge} V_\Sigma,$
$\mathbf{t}$: implicitly defined by $\theta^0 = i_{\mathbf{t}} g$.

$SO(3)(M) \leftarrow SO(2)(M)$:      At this step, $G/H = S^2$, whose first non-trivial homotopy group is in two dimensions. The topological obstruction for the reduction of $SO(3)(M)$ to $SO(2)(M)$ then becomes a 3-cocycle:

$$\mathscr{a}_3 {\in} H^3(M, \mathbb{Z}).$$

The fundamental tensor field for this reduction is a spacelike unit vector field $\mathbf{n}$ on M. As discussed in other work by the author [**21, 22**], this reduction is crucial to the appearance of wave motion on M, since the spacelike unit vector vector field that one must define is normal to the isophase submanifolds. Note that, in order to get this far, we have defined not just two *vector fields*, but a global 2-*frame field* $\{\mathbf{t}, \mathbf{n}\}$. Hence, the $\mathbb{Z}_2$-reduction of $\mathscr{a}_3$ is easily seen to be $w_3$, the third Stiefel-Whitney class of T(M). Furthermore, the homotopy classes of 2-frame fields are parameterized by $H^2(M, \mathbb{Z})$.

Insofar as we have defined a global timelike orthonormal 2-frame field on M in order to reduce from SL(4) to SO(2) as a structure group, we have also defined a rank-2 timelike sub-bundle B(M) of T(M) by way of the subspaces that are spanned by the frame $\{\mathbf{t}, \mathbf{n}\}$ at each point of M.  In fact, since we have a pseudo-metric at hand, we can also define a second rank-2 spacelike sub-bundle $\Phi$(M) of T(M) in the form of the orthogonal complement to B(M); hence, there is another Whitney splitting of T(M):

$$T(M) = B(M) \oplus \Phi(M).$$

At this point, let us make an important detour.  It happens (cf., Steenrod [**16**]) that the existence of a global rank-2 sub-bundle of T(M) is equivalent to the existence of a pseudo-metric of signature $(-1, -1, +1, +1)$.  This would be the fundamental tensor field that we would associate with making the reduction from SL(4) to SO(2, 2), a subgroup that has more structure to it than SO(2) as a group.  It not only contains the expectable SO(2)×SO(2) subgroups as maximal torii, which follows from the fact that the pseudo-metric gives the overall appearance of the difference between two two-dimensional Euclidian metrics, but it also contains a four-dimensional submanifold of strains that makes SO(2, 2) six-dimensional and non-abelian.  However, although the group SO(2, 2) has more structure than SO(2), that fact is only a consequence of the fact that we have not made any constraints on the rank-2 sub-bundle that defined the reduction to SO(2, 2), as we did in the process of reducing to SO(2) through a *chain* of subgroups.  Now, the reduction to a spacelike rank-2 vector sub-bundle of T(M) represents the differential system on spacetime that gives us the foliation of the proper time simultaneity leaves by momentary wave fronts, so we need to keep the physics of waves firmly in mind.  Hence, we pass over the reduction to SO(2,2) as being less physically motivated than the reduction to SO(2) that we defined.  The group SO(2, 2) might, however, still be of interest in the study of interacting fields with an SO(2) internal symmetry.

The introduction of the unit spacelike vector field $\mathbf{n}$ allows us to define a number of subsidiary geometrical objects:

A unit spacelike 1-form:  $\theta^1 = i_{\mathbf{n}}g = i_{\mathbf{n}}g_\Sigma$

A Lorentz metric on B(M):  $g_B = \theta^0 \otimes \theta^0 - \theta^1 \otimes \theta^1$

A volume element on B(M):  $V_B = \theta^0 \wedge \theta^1$

A Riemannian metric on $\Phi$(M):  $g_\Phi = g_\Sigma - \theta^1 \otimes \theta^1 = \theta^0 \otimes \theta^0 - \theta^1 \otimes \theta^1 - g$

A volume element on $\Phi$(M):  $V_\Phi = i_{\mathbf{n}}V_\Sigma = i_{\mathbf{n}^\wedge \mathbf{t}}V$

Conversely, we can use the objects $\theta^1$, $g_B$, $V_B$, $g_\Phi$, $V_\Phi$, that were introduced at this level of the chain to define:

$g_\Sigma = \theta^1 \otimes \theta^1 + g_\Phi$

$V_\Sigma = \theta^1 \wedge V_\Phi,$

$g \ = g_B - g_\Phi$

$V \ = V_B \wedge V_\Phi.$

$\mathbf{n}$:  implicitly defined by $\theta^1 = i_{\mathbf{n}}g_B$

$\theta^0 = i_{\mathbf{n}}V_B$

$\mathbf{t}$:  implicitly defined by $\theta^0 = i_{\mathbf{t}}g_B.$

SO(2)(M) ← $\mathbb{Z}_2$(M): At this stage, G/H = $\mathbb{R}P^2$, which has its first non-trivial homotopy group in dimension 1, namely, $\mathbb{Z}_2$.  There is no fundamental tensor field for this reduction, but there is yet another line field that we have defined on M, this time a spacelike one P(M) in the two-dimensional tangent subspaces of $\Phi$(M). The topological obstruction is then a cocycle:

$$\partial_2 \in H^2(M, \mathbb{Z}_2)$$

that represents $w_2$, the $2^{nd}$ Stiefel-Whitney class of M, and the elements of $H^1(M, \mathbb{Z}_2)$ tell us about the possible homotopy classes of P(M). In particular, if M is simply connected then all such line bundles will be homotopic.

$\mathbb{Z}_2(M) \leftarrow \{e\}(M)$: Finally, we address the issue of orienting the spacelike line bundle P(M). This is still non-trivial since G/H $= \mathbb{Z}_2$, hence, $\pi_0$(G/H) $= \mathbb{Z}_2$ is non-trivial. The topological obstruction to this step is then a 1-cocycle:

$$\partial_1 \in H^1(M, \mathbb{Z}_2),$$

which is the $1^{st}$ Stiefel-Whitney class of P(M). The fundamental tensor field for this step is a spacelike non-zero vector field **p** that generates P(M). It represents a global choice of zero phase for the vectors in each two-dimensional spacelike tangent subspace that is orthogonal to the one spanned by {**t**, **n**}. The homotopy class of **p** is determined by $H^0(M, \mathbb{Z}_2)$, as usual, so if M is path-connected there are two homotopy classes for **p**.

Notice that by the time we have reached this point, we have defined a global 3-frame field {**t**, **n**, **p**}, and, as a consequence, a global 4-frame field, {**t**, **n**, **p**, **q**}, since there is a unique spacelike line bundle that is orthogonal to the tangent subspaces spanned by{**t**, **n**, **p**} and since we have already oriented GL(M) there is a unique non-zero vector field **q** that generates it. In other words, in order for us to achieve this last reduction M would have to be *completely parallelizable*. This is consistent with the fact that in order to reach this step all of the Stiefel-Whitney classes would have to vanish, but this is a necessary condition that is not, however, sufficient.

If we note that the 3-frame field {**t**, **n**, **p**} spans a rank-3 sub-bundle of T(M), we see that, from the theorem in Steenrod [**16**] cited before, the existence of such a sub-bundle should be topologically equivalent to the existence of a pseudo-metric of type (+1, −1, −1, −1), but this is what we already got from the existence of a line bundle, up to an overall negative sign. This points to the fact that the existence of a Lorentz metric of type (−1, +1, +1, +1), which corresponds to a timelike line bundle, and a Lorentz metric of the type (+1, −1, −1, −1), which corresponds to a rank-3 spacelike sub-bundle, also defines a topological equivalence. Indeed, this topological equivalence also derives from the more general diffeomorphism of the Grassmann manifold $V_{1,n}$ with $V_{n-1,n}$.

Having reached the ultimate level of reduction, one must note that all of the geometrical fields that were defined on the way down can be constructed in terms of the global coframe field $\theta^i$, $i$ = 0, 1, 2, 3, where $\theta^3 = i_{\mathbf{q}}g$:

$$g_B = \theta^0 \otimes \theta^0 - \theta^1 \otimes \theta^1,$$
$$g_\Phi = \theta^2 \otimes \theta^2 + \theta^3 \otimes \theta^3,$$

$$g = \theta^0 \otimes \theta^0 - \theta^1 \otimes \theta^1 - \theta^2 \otimes \theta^2 - \theta^3 \otimes \theta^3,$$
$$V_B = \theta^0 \wedge \theta^1,$$
$$V_\Phi = \theta^2 \wedge \theta^3,$$
$$V = \theta^0 \wedge \theta^1 \wedge \theta^2 \wedge \theta^3,$$
$$\{\mathbf{t}, \mathbf{n}, \mathbf{p}, \mathbf{q}\} = \text{metric dual to } \{\theta^0, \theta^1, \theta^2, \theta^3\}.$$

**4. The singularity complex.** As mentioned in the introduction, the scope of the foregoing discussion subsumes the picture that was described for ordered media in the introductory remarks when one works under the assumption that one is dealing with a trivial G-principal bundle over M, i.e., G×M → M. As pointed out before, the various non-contractible homogeneous spaces G/H that we defined along the way can be though of as phases of the spacetime vacuum manifold. The *order parameter* in each phase is then what we have been calling "the equivariant map from G(M) to G/H that is defined by the reduction to H(M)," "the section of G/H(M) that defines the reduction," or "the fundamental tensor field of the reduction," when it exists. Furthermore, we shall treat the transformations of A(4) that produce contractible homogeneous spaces as deformations of the aforementioned phases, or excitations of the vacuum ground state in that phase, which would be any choice of H-reduction such that G/H is not contractible.

Notice that we are dealing with several examples in which the fact that the fundamental tensor field takes its values in a vector space does not imply that we are considering a *linear* field theory, since the range of values it can take is generally confined to a nonlinear manifold that only happens to be embedded in that vector space. For instance, a unit vector field takes its values in a sphere and the spacetime metric tensor takes its values in the homogeneous space SL(4)/SO₀(3,1), at least as we discussed it here ([6]).

However, since we are not generally assuming that M is completely parallelizable − hence, GL(M) is not trivializable − it is not prudent to assume that the reductions are trivializable either. Consequently, one must not only look at the homotopy groups of G/H and the homotopy classes of maps from M to G/H, as in condensed matter, but at the homotopy classes of global sections of the associated homogeneous bundle G/H(M) = G(M)×_G G/H. The relevant cocycles are then the ones that we discussed, which are cocycles in H*(M; π₊(G/H)). Again, for the trivial case, the associated homogeneous bundle becomes simply M×G/H, and its sections are maps from M to G/H. For the particular chain of subgroups of A(4) that we chose, the cohomology classes that played a role were the Euler class $e[M]$ and the Stiefel-Whitney classes $w_i[M]$ of T(M) − or GL(M), if you prefer.

In order to give a little more physical sense to the cocycles that we encountered, note that when M is orientable and oriented, the *k*-cocycles correspond to 4−*k*-cycles by Poincaré-Alexander duality. For instance, the Euler class, which is an element of H⁴(M), corresponds to a 0-cycle, i.e., set of points with associated integer coefficients. This is consistent with the fact that the Poincaré-Hopf theorem says that the obstruction to the existence of a global non-zero vector field **v** on M – i.e., a global 1-*frame field* – is going to be a set of isolated points with associated indices for **v**, at which **v** goes to zero. The formal sum over the points, weighted by the indices, gives a 0-chain, which is automatically closed, hence, a 0-cycle in the integer homology of M. By duality, it corresponds to a 4-cocycle, which when it is evaluated on the fundamental cycle of the orientation on T(M), gives the Euler-Poincaré characteristic.

In fact, in Stiefel's paper on the parallelizability of manifolds the obstructions that he defined were $\mathbb{Z}_2$-cycles that generalized this scenario ([7]). He referred to the set of points at which a $k$-frame field on an $n$-manifold is not defined, when suitably triangulated, as the "singularity complex," and his innovation was to show that it was a $\mathbb{Z}_2$-cycle, i.e., that when the coefficients of the simplexes from which it was constructed were replaced with 0 or 1 according to whether they were even or odd, respectively, then the resulting chain would be closed.

Our immediate challenge is to relate the usual terminology of topological defects, which are elements of $\pi_*(G/H)$, to the cocycles of H*(M; $\pi_{*-1}(G/H)$) in a manner that seems physically intuitive; fortunately, this is not as difficult as it sounds. The key to proceeding is seeing that one first generalizes from homotopy classes of maps from $k$-spheres, which presumably represent a highly symmetric form of space or spacetime, into G/H to homotopy classes of maps from $k$-spheres that are defined in the $k$+1-cells of the $k$+1-skeleton of M – when M is represented homotopically as a CW-complex – to G/H; these will still represent elements of $\pi_k(G/H)$, but in a local form. If the type of a local topological defect – wall, string, monopole, texture – is determined by the first non-vanishing $\pi_k(G/H)$ then what the singularity complex for a section of G/H(M) represents is a powerful generalization of the networks of local defects that are currently being considered by physics. A network represents the 1-dimensional case of a chain complex, which means that if the network is a 1-cycle then it is Poincaré dual to a 3-cocycle.

Presumably, the cells of the singularity complex should also represent the "sources" of the fundamental fields that are associated with the various reductions. The *charge* that we associate with a topological defect is then the corresponding element of $\pi_k(G/H)$.

For the seminal example of Poincaré-Hopf, the singularity complex of a vector field **v** is a 0-chain that consists of the points $p$ at which $\mathbf{v}(p) = 0$, and the associated homotopy class of Gauss maps with the same winding number $N_p$ for some sufficiently small $n$−1-sphere surrounding it, i.e., an element of $\pi_{n-1}(S^{n-1}) = \mathbb{Z}$. (The winding numbers in question are also determined by the choice of **v**.) When we form the sum:

$$\sum_p N_p \, p$$

<hr>

[7] Not surprisingly, Stiefel was a student of Hopf who was trying to generalize the scope of the theorem that his mentor had established.

we have formed a 0-chain with values in $\pi_{n-1}(S^{n-1})$, i.e., an element of $H_0(M; \mathbb{Z})$. By duality, this gives an element of $H^n(M; \mathbb{Z})$, which happens to be the Euler class of T(M).

In the language of topological defects, one could say that an element of $H_0(M; \pi_{n-1}(S^{n-1}))$ *associates* a wall, string, monopole, or texture defect (depending on *n*) *with each singular point p*. For example, if M is two-dimensional, we are saying that Poincaré-Hopf associates a zero of any vector field on M with a string defect. If one considers the zeroes of a vector field to be fixed points of its associated local flow then these defects can represent vortices. The generalization to higher-dimensional elements of the singularity complex is straightforward. Dually, we could say that an obstruction cocycle in dimension *k* associates such a topological defect (in dimension *k*-1) with each *k*-cell in *the complex that represents* M. This would put the vortices of a vector field on $S^2$ into the complement of the zeroes. The actual value in $\pi_{n-1}(S^{n-1})$ that one associates with the topological defect is essentially a "charge." In the two-dimensional case, the Euler-Poincaré characteristic then represents the vortex strength in the flow about the singular point.

We shall now go over the steps in our chain of reductions of A(M) at which obstructions appeared and interpret them in the language of topological defects.

GL(M) → GL⁺(M): The obstruction to the orientation of M is the first Stiefel-Whitney class of T(M), which is an element $w_1[M] \in H^1(M; \pi_0(GL(4)/GL^+(4))) = H^1(M; \mathbb{Z}_2)$. Hence, by Poincaré-Alexander duality, it corresponds to a 3-cycle in $H_3(M; \mathbb{Z}_2)$, i.e., a formal sum of closed 3-cells whose coefficients are either +1 or –1. What we have done, in the language of topological defects, is to locally associate a homotopy class of maps $[\varnothing ; GL(n)/GL^+(n))]$, i.e., wall defect, with each 3-cell in the formal sum that represents the singularity complex for the orientation. The +1 or –1 then represents a choice of "side" for the wall. Although a fundamental tensor field – the volume element – does not show up until the next reduction – to SL(4) – since that reduction does not change the homotopy type of the reduction, we take the liberty of associating the wall defect with the volume element.

SL(M) → SO(3,1)(M): The obstruction to defining a Lorentz metric on M, at least in the case of compact manifolds, is the Euler class $e[M] \in H^4(M; \pi_3(\mathbb{R}P^3)) = H^4(M; \mathbb{Z})$. From our previous discussion of the Euler class, this represents the association of a texture defect to each point of the singularity complex for the metric tensor field, which gives a 0-cycle. The fundamental tensor field is, of course, the Lorentz metric tensor field, which we see is associated with texture defects. Since one also associates mass distributions in space as the sources of gravitational fields, this suggests a pointlike nature to elementary masses in space, i.e., worldlines in spacetime.

SO(3,1)(M) → SO₀(3,1)(M): Once again, we are looking at a problem of orientation, only this time we are orienting a different vector bundle, so we are still associating a wall defect with each 3-cell in the singularity complex for the time orientation **t**. However, this does not have to be the same as the singularity complex for the orientation of T(M).

SO(3)(M) → SO(2)(M):    The obstruction to the reduction from spacelike orthonormal 3-frames to spacelike orthonormal 2-frames is an element of $H^3(M; \pi_2(S^2)) = H^3(M; \mathbb{Z})$. In terms of cohomology, this represents the association of a monopole defect to each 3-cell in the complex for the two-frame field that is associated with this reduction, namely {**t**, **n**}. Whether one chooses to regard the monopoles as the sources of the vector field **n** or the two-frame field {**t**, **n**} is moot.

SO(2)(M) ← $\mathbb{Z}_2$(M):  The obstruction to this reduction, which represents the choice of a "zero-phase" line P(M) for the spacelike two-dimensional subspace that is orthogonal to the subspace spanned by {**t**, **n**} at each point, is an element of $H^2(M; \pi_1(SO(2)/\mathbb{Z}_2) = H^2(M, \mathbb{R}P^2) = H^2(M, \mathbb{Z}_2)$, namely, $w_2[M]$. Hence, we are associating a string defect with every 2-cell in the singularity complex of the spacelike line field we are defining. One might say these 2-cells contained "vortices." Intuitively, one imagines that a consistent choice of zero phase for a vortex about the origin will become undefined at the origin. These vortices then serve as the source of the line field P(M).

$\mathbb{Z}_2$(M) ← {$e$}(M):   This is one last problem of orientation, so there is one more set of wall defects associated with the 3-cells of the singularity complex for the non-zero spacelike vector field **p** that we are defining. Again, this does not have to be the same set of 3-cells that appeared in either the choice of orientation of T(M) or the orientation of L(M). This wall defect is to be regarded as the source of the vector field **p**.

Now, let us drop the assumption that any of these obstructions necessarily vanish. Start at the top of the chain of subgroups that we have been considering and note that every time one encounters an obstruction in the form of a $k$-chain $c_k$ with appropriate coefficients the reduction at this step will always be defined on the complement of the subset of points in M that the $k$-chain represents. Hence, one can proceed down the chain by restriction until another obstruction is reached, restrict to the complement (within the first complement), and so on. Ultimately, one will reach an {$e$}-structure, i.e., an orthonormal frame field that is defined on the complement of the union of all of the subsets that are represented by obstruction cycles. If it were not for the fact that not all of the coefficients are necessarily in the same group, we could characterize this subset as the one that is represented by the sum of all of the obstruction cycles. However, since the only homotopy groups that we encountered in the present chain of subgroups were {0}, $\mathbb{Z}_2$, and $\mathbb{Z}$, we could form the sum of the $\mathbb{Z}_2$-reductions of the obstruction cycles, which will be dual to what one usually calls the *total Stiefel-Whitney class* of T(M) (minus one):

$$w[M] = 1 + w_1 + w_2 + w_3 + w_4.$$

One should not be deluded by the simplicity of its expression, since the individual terms represent $\mathbb{Z}_2$-cocycles of increasing dimensions, and there is nothing to say these cocycles do not represent sums with a large number of terms. For instance, a vanishing

Euler class may still represent a large number of zeroes and indices for some particular vector field.

## 5. Spacetime vacuum phase transitions.

Now that we have walked through a lot of steps that the mathematicians would regard as "straightforward, but tedious" (whereas the physicists would call them "too abstract to be physically important!"), we should go over them in the context of how the geometry and topology of spacetime that we just detailed relates to corresponding physical phenomena. We shall attempt to use wave mechanics as a unifying theme.

Certainly, of all of the notions that we encountered in the form of fundamental tensor fields defined by the various reductions in the sequence we have been considering, the notion of a spacetime metric has certainly been discussed more than the other ones. However, one must never lose sight of the fact that the geometry of Minkowski space and Lorentz manifolds are essentially abstracted from the structure of the characteristic equation for the linear wave equation that one derives from Maxwell's equations. Hence, since many physicists have been considering nonlinearity, especially as it relates to wave equations, as a fruitful direction into which one should proceed, we shall take the position that the appearance of a metric tensor field is subsidiary to the process of facilitating wave motion.

Even though the structure that was seen to be necessary for the definition of wave motion in its most general sense, namely, an SO(2)-structure, is several reductions down from the bundle A(M) that we started with, nevertheless, we shall try to give the reductions that preceded it a wave-theoretical form by treating the spacetime vacuum manifold as something like a continuum of points with associated phases that correspond to the various types of motion that it can support. Moreover, the transitions between phases are associated with the appearance of topological defects. In any given phase, the spacetime manifold is also subject to deformations that do not change its topological type, at least up to homotopy.

An important point to emphasize is that in order to account for the possibility that the points of the spacetime vacuum manifold can have various phases it is necessary to consider the spacetime manifold to be GL(M) instead of M itself. This is a natural extension of the notion of Cosserat continuum [**23, 24**], which pertains to the case of orthonormal frames in space.

Let us start with the most general case – next to A(M) – the case of motion in GL(M) for a possibly non-orientable M; indeed, we can regard A(M) as a sort of deformation of GL(M) by way of translations. Notice that non-orientability would not preclude the existence of an everywhere non-zero vector field with a global flow on M; one can easily imagine such a flow on a Möbius band. The only consequence of non-orientability in such a case is that if one followed a local area element around the band, unless it contracted to zero at some point, it would become inconsistent with itself after one complete loop. Hence, one could not define an *incompressible* flow on a non-orientable

M. We refer to the phase of the spacetime vacuum manifold after the internal symmetry group has broken to SL(4) as the *incompressible phase.* This phase is subject to deformation by dilatations.

It is interesting to note that before one can define wave motion, one must first satisfy an incompressibility constraint, since the counter example of acoustics would seem to contradict this on the surface of things; however, the propagation of waves in compressible media is generally associated with a small-displacement approximation, so this is not really a contradiction. The response of a compressible medium, such as air, to large enough driving displacements would be corrupted by such things as turbulence. Nevertheless, one should not lose sight of the fact that we are presumably dealing with the phases of the spacetime *vacuum.* Hence, an SL(4)-reduction of A(M) represents a vacuum ground state in the incompressible phase and the dilatations define its excitations.

The transition from the compressible phase to an incompressible phase is associated with a topological defect in the form of a wall that one associates with every 3-cell in the singularity complex for the reduced bundle. The corresponding charge on the wall is simply the + sign that represents a choice of side. In $\mathbb{Z}_2$-cohomology this defect is represented by the first Stiefel-Whitney class for T(M), $w_1[\mathrm{M}]$.

Defining a metric on spacetime facilitates wave motion by defining a notion of causality, in the form of the light cones in each tangent space. In the theory of hyperbolic quasilinear second order PDE's, these cones are characteristic manifolds, and relate to the question of whether the Cauchy problem for a given spacelike three-dimensional submanifold is well posed. Interestingly, any timelike line $\mathrm{L}_x(\mathrm{M})$ at each $x{\in}\mathrm{M}$ will generate the same pair of light cones ([8]), so although defining a Lorentz metric is equivalent to defining such a timelike line, the choice is not canonical. On a compact spacetime, the obstruction to the construction of such a global line field L(M), when viewed as a set of topological defects that are generated by the transition from the incompressible phase to the *Lorentz phase*, associates a texture defect to each 0-cell, i.e., point, in the singularity complex of the reduction. This obstruction is simply the Euler class of T(M); its charge is the Euler-Poincaré characteristic of M. This phase of the spacetime vacuum manifold is subject to deformations by means of dilatations and Lorentz shears.

Since the light cones consist of two nappes, either of which could represent motion into the future or past for tangent vectors at each point, a further step in the direction of wave motion is to eliminate the ambiguity by choosing a time orientation. The choice of timelike vector field **t** that makes this possible also defines a notion of *proper time* and a *rest frame* at each point. Actually, a rest frame at a point is an SO(3)-equivalence class of Lorentz orthonormal 4-frames that share **t** as their common timelike element; that is, any such orthonormal 4-frames that differ by a spacelike rotation are just as acceptable for rest frames.  This is also why the reduction from an $SO_0(3,1)$-structure on spacetime to an $SO(3)$-structure is immediate: one simply reduces to the $\mathbf{t}$-adapted frames (i.e., rest frames) in each such equivalence class.  We shall think of this phase as the *time-oriented Lorentz phase* of the spacetime vacuum manifold.  Since the homogeneous space $SO_0(3,1)/SO(3)$ is contractible, and represents the boost transformations that can be applied to the rest frame, we shall treat the non-rest frames as deformations of rest frames by way of boosts.  In effect the rest frames represent the vacuum ground states in this phase and the boosts represent excitations.

---

[8] In fact, a global choice of light cone in each tangent space is associated with the reduction to a CO(3,1)-structure, since the light cones are invariant under conformal Lorentz transformations.

The transition from the Lorentz phase to the time-oriented Lorentz phase is associated with a topological defect that takes the form of associating a wall defect to every 3-cell in the singularity complex of the reduction; this corresponds to $w_1[L]$ in $\mathbb{Z}_2$-cohomology. Again, the associated charge is $\pm 1$.

So far, we still have yet to define a particular example of wave motion in spacetime. What is left is to associate the points of each rest-space with each other in the eyes of "phase."  The way that we do this is to choose a spacelike non-zero vector field $\mathbf{n}$ that will represent the normal to the spacelike isophase surface through each point.  Together with $\mathbf{t}$, it defines a timelike Lorentz-orthogonal 2-frame that spans a 2-plane at each point, which is, in turn, the orthogonal complement to a spacelike two-dimensional vector sub-bundle $\Phi(M)$ of $T(M)$; i.e., a differential system on M.  We call this phase the *wave phase* of the spacetime vacuum manifold.  Any choice of F(M) represents a vacuum ground state for the wave phase and the degeneracy of the vacuum ground state is indexed by the homotopy classes of all possible $\Phi(M)$.

The transition from the time-oriented Lorentz phase to the wave phase is associated with a set of topological defects that associate monopoles with each 1-cell in the singularity complex of the reduction.  The associated charge is an integer that represents a winding number for a sufficiently small 2-sphere about the monopole.  In $\mathbb{Z}_2$-cohomology this set of defects is represented by $w_3[M]$.

Since $\mathbf{t}$ was really an arbitrary choice of proper time line or rest frame, the differential system $\Phi(M)$ should be indifferent to any local Lorentz transformations of the 2-frame $\{\mathbf{t}, \mathbf{n}\}$, which would take the form of boosts.  This represents one sort of "gauge equivalence" defined on GL(M).  Another is the $SO(2)$ gauge invariance that arises from the fact that we still have not chosen an actual spacelike 2-frame to complement $\{\mathbf{t}, \mathbf{n}\}$, i.e., to frame $\Phi(M)$.  It is worth pointing out that $SO(2)$ gauge invariance plays a fundamental role in two of the immediate physical examples of wave theories, namely, quantum wave mechanics and electromagnetism.  However, one must also keep in mind that there are solutions to Maxwell's equations that are not wavelike, e.g., static solutions.

The $SO(2)$ gauge equivalence is broken to $\mathbb{Z}_2$ by making a choice of spacelike zero-phase axis P(M) in the two-dimensional spacelike isophase subspaces at each point.  We call this phase the $SO(2)$-*gauged phase* of the spacetime manifold.  The transition from the wavelike phase to the $SO(2)$-gauged phase is associated with a set of topological defects in the form of strings associated with each 2-cell in the singularity complex of the reduction; the charge on each string is ±1. In $\mathbb{Z}_2$-cohomology this set of defects is represented by $w_2[M]$.

This $\mathbb{Z}_2$ symmetry can be broken to the identity group by choosing a spacelike nonzero vector field **p** that takes its values in P(M), since the effect of making such a choice is to imply the existence of one last spacelike vector field that completes the orthogonal tetrad at each points in the complement of the total singularity complex for this last reduction. In effect, parallelism destroys the wavelike character of the motion that was defined up to that step in the chain of reductions. We shall call this final phase the *completely parallelized phase* of the spacetime manifold. It represents an ultimate phase in that there are no further reductions possible, so, in a sense, we can regard the completely parallelized phase as a vacuum ground state. The degeneracy in the vacuum ground state of the completely parallelized phase relates to the homotopy classes of maps from M into $\mathbb{Z}_2$. If M is path-connected then there will be only two such classes, which correspond to the left and right orientations for the frames.

Conversely, we can start with a ground state in the completely parallelized phase and regard all of the other states that were described up to now as the result of excitations that take one of two forms: deformations within phases and phase transitions. This is quite consistent with the usual picture in condensed matter, as well as the usual approach to the appearance of self-organization is systems far from equilibrium versus the effects of small perturbations. Indeed, it suggests a realm of quantum fluctuations about the vacuum ground state beyond those amenable to perturbative analysis in which phase transitions take place; of course, this concept is not new in quantum field theory.

The transition from the SO(2)-gauged phase to the completely parallelized phase is associated with a set of topological defects in the form of walls associated with the 0-cells in the singularity complex of $\Phi(M)$ and associated charges of ±1; this situation is represented in $\mathbb{Z}_2$-cohomology by $w_1[\Phi]$.

**6. Discussion.** The ulterior motive of this study was to perform the actual calculations of the topological obstructions that were associated with a chain of reductions of GL(M) that happens to be of particular relevance to the problem of defining the intrinsic geometrical and topological character of wave motion, and independently of the wave equation that one chooses to represent it. Hence, we will first summarize the geometrical and topological nature of the wave phase, and then point out some further topics to be addressed that will bring this work closer to the work that was described in [**8,9**].

To summarize the nature of the wave phase of the spacetime vacuum manifold:

 *a)* The topological obstructions that had to vanish were $w_1[M]$, $w_3[M]$, and $w_4[M]$. Hence, different expressions for $w_2[M]$ will correspond to topologically distinct wave phases.

*b)* The fundamental tensor fields that we have introduced in order to get this far have been a volume 4-form V, a Lorentz metric *g*, and a timelike orthonormal 2-frame {$\mathbf{e}_0 = \mathbf{t}$, $\mathbf{e}_1 = \mathbf{n}$}.

In the Part II of this study, we will address the corresponding geometrical aspects of the chain of reductions that was discussed in this work. In particular, we will look at the reductions and deformations of an affine connection to the various G-structures in the chain and try to give the resulting geometrical objects some sort of physically intuitive sense, especially in the context of wave motion. As was pointed out above, although some of the reductions in the chain of subgroups produce no change in the homotopy type of the spacetime vacuum manifold, they can nevertheless produce useful geometric information. We shall then give more attention to this aspect of the reduction process in the sequel.

Beyond the aforementioned level of analysis, as [**9**] discussed at length, if wave motion is to generate a transversal pair of – possibly singular – codimension-one foliations of spacetime, namely the simultaneity foliation and the isophase foliation, then one must address the question of the integrability of the sub-bundle $\Phi(M)$. In the context of differential systems, we are asking whether the restriction of the spacelike two-dimensional isophase sub-bundle $\Phi(M)$ to a simultaneity leaf $\mathscr{L}$, which, in turn can be regarded as a differential system $\Phi(\mathscr{L})$ on $\mathscr{L}$, is the tangent bundle to the two-dimensional spacelike leaves of a codimension-one foliation of $\mathscr{L}$ that can be obtained by intersecting the three-dimensional timelike isophase leaves with $\mathscr{L}$. In the context of G-structures, we are asking whether the SO(2)-frame bundle that we have defined can be generated by the natural frame fields that are associated with the coordinate charts of some two-dimensional atlas for each leaf. The integrability of G-structures is a deep subject that rapidly leads into further complexities, such as Spencer cohomology, although it is getting increasing attention in the context of continuum mechanics [**25**] and the geometry of PDE's [**26**]. The integrability of spacetime G-structures will define the objective of Part III of this study.

# Appendix: CW-complexes

Although the literature of CW-complexes ([9]) in algebraic topology [27-30] is vast and potentially abstruse, that is only because the representation of a topological space (up to homotopy equivalence) as a CW-complex is that convenient to many applications. In fact, the basic notions are not difficult to discuss. Hence, we shall only briefly touch upon the basics as they relate to the subject at hand, namely, spacetime topological defects.

The essential process is one of constructing the topology of a space, up to homotopy equivalence, by means of a set of elementary cells of various dimensions that are attached to the boundaries of the next higher-dimensional cells. Finite complexes, i.e., ones with a finite number of cells, will represent compact spaces, so in order to represent a non-compact space as a CW-complex, one will necessarily need an infinitude of them.

A *k-cell* $E_k$ in a topological space M is the homeomorphic image of either an open *k*-ball or an open *k*-cube $I^k$, depending on one's intended application. (Of course, they are mutually homeomorphic, anyway; indeed, they are both homeomorphic to $\mathbb{R}^k$.)

A common way of constructing topological spaces from more elementary ones is the *attaching construction.* In order to attach the topological space X to the space Y, one needs to find subsets $A \subset X$ and $B \subset Y$ that are homeomorphic by some map $f$: $A \to B$. One then defines an equivalence relation on the disjoint union $X \vee Y$ by $x \sim y$ iff $f(x) = y$. When one passes to the quotient $X \vee Y \to X \vee Y/\sim$ the resulting space is said to result from "attaching X to Y at A by way of $f$." For most topological purposes, the choice of A, B, and $f$ is inconsequential, i.e., any other choice would give a different space that is nonetheless homeomorphic to the one that one obtained from the first choice. A simple example of attaching two spaces is choose an arbitrary point $a$ in the circle $S^1$ and attach two copies of $S^1$ at $a$ by way of the identity map; one obtains a figure-eight space. Slightly less elementary is the way that attaching two closed solid *n*-torii at their boundaries by way of the identity map produces an open ball.

When the attaching construction is applied to *k*-cells, it leads to the notion of CW-*complexes*, which are topological spaces that one obtains by starting with a set points and beings attaching *cells* of higher dimension at their boundaries, by which, we mean the points of $\overline{E}_k - E_k$. When the set of all such cells is finite, one obtains a compact topological space; otherwise, the resulting space *can* be non-compact. In particular, a finite CW-complex is a compact topological space X and a sequence of closed subspaces:
$$X^0 \subseteq X^1 \subseteq \ldots \subseteq X^n = X,$$

such that $X^0$ is a finite set of points and for each $0 < k \leq n$ the subspace $X^k$ is obtained by attaching a finite number of *k*-cells to $X^{k-1}$ at subsets of their boundaries; one refers to each $X^k$ as the *k-skeleton* of X.   This has the effect of making each $X^k - X^{k-1}$ homeomorphic to the disjoint union of a finite number of *k*-cells.  Since one also has that

$$X = \bigcup_{k=1}^{n}(X_k - X_{k-1}),$$

this says that the space X is partitioned by the interiors of the cells.  In order to resolve the possible confusion about why this does not make X disconnected with all of the cells as components, we need to add that the topology that one gives X in terms of this decomposition – which is called the *weak* topology – does not make all of the cells open sets. A subset of X is closed in the weak topology iff its intersection with all of the cells in the partition is closed in the original topology of X.  (Keep in mind that the cells do not all have the same dimension, so one would not expect each of them to contain sufficiently small *n*-balls about each point.)

---

[9] The origin of the C and the W in the term is that the C comes from "closure finite" and W, from "weak topology." A CW-complex has nothing to do with the paranoid delusion that one is really a famous Country and Western star.

For the purposes of this article, we shall refer to a set of cells that partition a topological space M and identifications that attach them together into a space with the homotopy type of M as a *realization* of M by a CW-complex, which is generally not unique, and we shall denote a realization of M by |M|.

One advantage of the CW-complex construction is that it makes computing the homology of X simpler once one has a – not generally unique – representation for X as a CW-complex.  First, one notes that:

$$H_j(X^k, X^{k-1}) = \begin{cases} 0 & \text{when } j \neq k \\ \text{free Abelian group with one generator for each } k\text{-cell} & \text{when } j = k. \end{cases}$$

One then defines a chain complex by the direct sum of the $C_k = H_k(X^k, X^{k-1})$ for $0 \leq k \leq n$, with $X^{-1} = \varnothing$, and a boundary operator that we will not discuss in detail here ([10]), and shows that, in fact the homology that one obtains from this differential graded module is, indeed, isomorphic to the singular homology of X.  Since the set of all singular *k*-chains in X is generally uncountable and the set of all *k*-cells is finite, this represents a considerable simplification.  Of course, one must first represent X as a CW-complex. Even then, not all CW-complexes will define topological manifolds.  However, it is a well-known result [17] that any compact differentiable manifold is homotopically equivalent to a finite CW-complex ([11]).  In fact, one can weaken that statement by requiring only the existence of a *Morse function,* i.e., a smooth function *f* with non-degenerate critical points, with the property that $f^{-1}(x \leq a)$ is compact for all $a \in \mathbb{R}$ ; in that case, the CW-complex might have infinitely many cells.

One examples of realizing a topological space by a CW-complex is to regard the *n*-sphere as an *n*-ball $E_n$ attached to a point $E_0$ by identifying the boundary of the ball to that point. One can also represent $S^n$ by two *n*-cells, which represent the upper and lower hemispheres, identified at their boundary; i.e., the equator.

---

[10] It is defined in terms of the "connecting homomorphism" of the triple $(X^k, X^{k-1}, X^{k-2})$; cf. [27-30].

[11] An obvious approach to achieving this decomposition is by Morse theory [32].

Realizations for the various real projective spaces can be obtained recursively by noting that $\mathbb{RP}^n$ is homotopically equivalent to $\mathbb{RP}^{n-1}$ with an $n$-cell attached. Hence, one will realize $\mathbb{RP}^n$ by a set of $n+1$ cells, one in each dimension (including zero). In fact, this result extends to $\mathbb{RP}^\infty$, which clearly produces an infinite CW-complex.

The realization of an $n$-torus is somewhat harder to visualize. One starts with a point, attaches $n$ 1-cells (open unit intervals) to that point at their boundaries, and thus produces the "wedge" of $n$ circles. One then attaches a single $n$-cell to that wedge of circles at its boundary to produce the $n$-torus.

The Stiefel manifold $V_{n,m}$ can be given a realization in terms of a finite CW-complex by means of *Schubert cells*. We shall simply direct the curious to the literature [**17**] for the details.